\documentclass{article}
\usepackage{ijcai21}

\usepackage{times}
\usepackage{soul}
\usepackage{url}
\usepackage[hidelinks]{hyperref}
\usepackage[utf8]{inputenc}
\usepackage[small]{caption}
\usepackage{graphicx}
\usepackage{amsmath}
\usepackage{amsthm}
\usepackage{booktabs}
\usepackage{algorithm}
\usepackage{algorithmic}
\newcommand*{\thead}[1]{%
\multicolumn{1}{c}{\begin{tabular}{@{}c@{}}#1\end{tabular}}}
\usepackage{tikz}
\usepackage{subcaption}
\usepackage{pgfplots}
\usepackage{amssymb}

\title{General-Purpose Speech Representation Learning through a Self-Supervised Multi-Granularity Framework}

\author{
Yucheng Zhao$^1$
\and
Dacheng Yin$^1$\and
Chong Luo$^2$\and
Zhiyuan Zhao$^2$\and
Chuanxin Tang$^2$\and
Wenjun Zeng$^2$\and
Zheng-Jun Zha$^1$
\affiliations
$^1$University of Science and Technology of China\\
$^2$Microsoft Research Asia
\emails
\{lnc, ydc\}@mail.ustc.edu.cn,
\{cluo, zhiyzh, chutan, wezeng\}@microsoft.com,
zhazj@ustc.edu.cn
}

\begin{document}

\maketitle

\begin{abstract}
This paper presents a self-supervised learning framework, named MGF, for general-purpose speech representation learning. In the design of MGF, speech hierarchy is taken into consideration. Specifically, we propose to use generative learning approaches to capture fine-grained information at small time scales and use discriminative learning approaches to distill coarse-grained or semantic information at large time scales. For phoneme-scale learning, we borrow idea from the masked language model but tailor it for the continuous speech signal by replacing classification loss with a contrastive loss. 
We corroborate our design by evaluating MGF representation on various downstream tasks, including phoneme classification, speaker classification, speech recognition, and emotion classification. 
Experiments verify that training at different time scales needs different training targets and loss functions, which in general complement each other and lead to a better performance.

\end{abstract}

\section{Introduction}

Unsupervised pre-training, or representation learning, has drawn wide interests in both academia and industry. The BERT model \cite{DBLP:conf/naacl/DevlinCLT19} has become a universal feature extractor for solving a wide range of natural language processing (NLP) tasks. Recently, it is reported that the image embedding learned in an unsupervised manner achieves comparable performance to its supervised counterparts in the image classification task \cite{DBLP:conf/cvpr/He0WXG20,DBLP:conf/icml/ChenK0H20}. Actually, most contemporary unsupervised pre-training methods adopt the self-supervised learning approach. We use these two terms interchangeably in this paper to refer to methods that do not need human annotation. 

In the speech domain, pre-training is not a new concept. The speaker recognition task depends heavily on the supervised pre-training step to obtain a good feature embedding. Recently, self-supervised learning is also used to pre-train dedicated models for automatic speech recognition (ASR) \cite{DBLP:conf/interspeech/SchneiderBCA19,DBLP:conf/iclr/BaevskiSA20,DBLP:conf/nips/BaevskiZMA20,DBLP:conf/icassp/LingLSK20}. In this work, however, we are not focusing on these task-oriented pre-training. Instead, we aim to pre-train a general-purpose feature extractor which embeds a speech signal into a feature representation that could be used for a variety of downstream speech tasks, in a way similar to how pre-trained language and image representations are used in their respective domains. 

\begin{figure}[t]
\centering
\includegraphics[width=0.9\columnwidth]{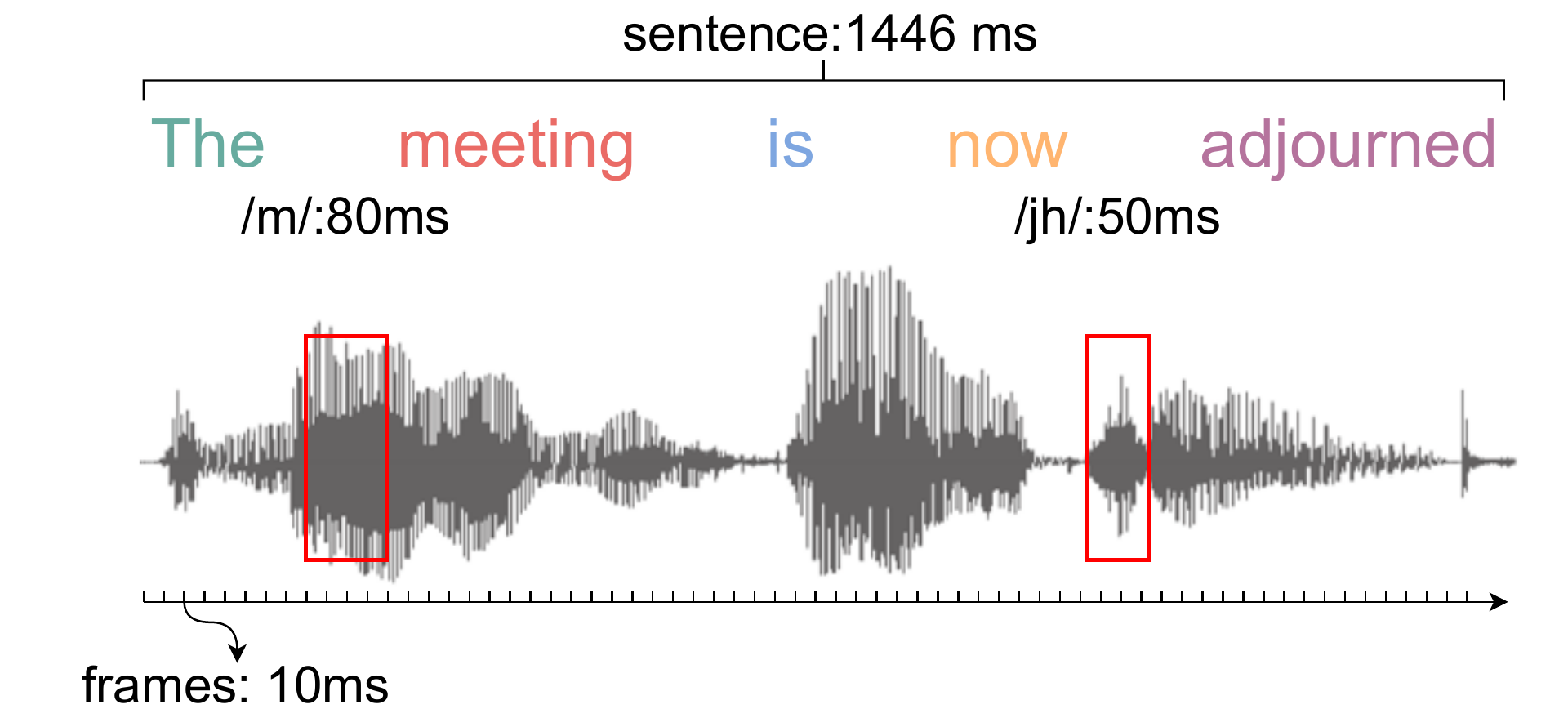}
\caption{Speech hierarchy. The waveform is sampled at 16K Hz. Sample points within a 10ms segment form a frame, which is the basic operating unit in many speech algorithms. Phonemic information can be extracted from several frames, as we illustrate with red boxes. The whole sentence lasts for more than one second. }
\label{fig:structure}
\end{figure}


The main difficulty in learning a general-purpose speech representation is that speech carries complex hierarchical structure (samples, phonemes, and sentences) which contains relevant information at different time scales \cite{DBLP:conf/interspeech/PascualRSBB19}. In this work, we propose a Multi-Granularity Framework, named MGF, to train the model at multiple time scales. A key innovation in MGF is to adopt different learning approaches for the learning at different time scales. In particular, we use generative approaches to capture fine-grained information for small time scales on the order of a few milliseconds, and we adopt discriminative approaches to distill semantic information for large time scales which correspond to a phoneme and a sentence. In order to realize phoneme-level contrastive learning, we extend the token-oriented masked language model (MLM) model \cite{DBLP:conf/naacl/DevlinCLT19} to continuous masked language model (cMLM) to accommodate the continuous speech signals without token boundaries. 
MGF is implemented by a deep bidirectional Transformer \cite{DBLP:conf/nips/VaswaniSPUJGKP17,DBLP:conf/naacl/DevlinCLT19}.

We evaluate the MGF representation on multiple downstream tasks and benchmark datasets, which becomes the second main contribution of our work. The performance of MGF is first evaluated on phoneme classification and speaker classification tasks, following the other general-purpose speech representation learning work \cite{DBLP:journals/corr/abs-1807-03748,DBLP:conf/icassp/LiuYCHL20}. 
We find that features learned by MGF is very powerful on these two orthogonal tasks. On the LibriSpeech dataset, MGF representation achieves a phoneme classification accuracy of 73.4\% under linear evaluation, surpassing the existing unsupervised pre-training methods by a large margin. On the speaker classification task, MGF representation is the first to achieve an accuracy of 100\%. 

We further evaluate MGF in other three downstream tasks. First, in view of the saturated performance in speaker classification, we propose a new and harder task named \textit{one-shot speaker classification}, where only one utterance per speaker is provided in the fine-tuning stage. In this task, MGF is evaluated against the well-known x-vector and d-vector and is shown to achieve better performance. Second, we compare MGF with a task-specific pre-training model wav2vec in the ASR task. Third, we test MGF representation on the IEMOCAP emotion classification task. Surprisingly, simply appending a fully-connected layer after MGF achieves the top performance among all existing audio-based approaches. 




\section{Related Work} \label{sec:2}

There are two camps of self-supervised learning approaches, namely discriminative and generative approaches. We will first review these two approaches for speech pre-training, and then discuss other related work that motivates MGF.

\subsection{Discriminative Approaches} \label{subsec:2.1}

Discriminative approaches acquire supervision signal from the contrastive distance between a selected positive sample and several negative samples. By carefully designing the training target and the data sampling procedure, samples can be automatically labelled. 


Contrastive predictive coding (CPC) \cite{DBLP:journals/corr/abs-1807-03748} is a contrastive learning method based on predicting the future in the latent space. 
The representations of temporally nearby segments are treated as positive samples while those of temporally distant segments are treated as negative samples. 
However, one could easily find a counter example in speech processing. For example, a word appears twice in an utterance with the same meaning. When the first appearance is the anchor, the second appearance should not be treated as a negative sample no matter how far it is. Previous work \cite{DBLP:conf/interspeech/ChungHTG19} also notices that the choice of negative samples in CPC has huge effect on its performance on the phoneme classification task.

While CPC itself is a general-purpose speech pre-training method, it can be leveraged in some task-specific pre-training models, such as wav2vec \cite{DBLP:conf/interspeech/SchneiderBCA19}, vq-wav2vec \cite{DBLP:conf/iclr/BaevskiSA20}, and wav2vec 2.0 \cite{DBLP:conf/nips/BaevskiZMA20}. Vq-wav2vec proposes a quantization algorithm so that wav2vec (which adopts CPC) can be combined with the BERT model \cite{DBLP:conf/naacl/DevlinCLT19} to achieve better performance. 
Wav2vec 2.0 improves vq-wav2vec by training the entire model end-to-end. It also uses a very large unlabelled dataset for pre-training. These task-specific pre-train models are very powerful in their target task, but perform poorly in other speech tasks. 

\subsection{Generative Approaches}

Generative approaches learn to reconstruct signal in the input space or features in some latent spaces. Training is supervised by the reconstruction loss. 
Autoregressive predictive coding (APC) \cite{DBLP:conf/interspeech/ChungHTG19} uses an autoregressive model to encode the history and predict the future. 
A follow-up work \cite{DBLP:conf/acl/ChungG20} adds an auxiliary objective which encourages the model to additionally remember the past. DeCoAR \cite{DBLP:conf/icassp/LingLSK20} borrows the bidirectional learning idea from ELMo \cite{DBLP:conf/naacl/PetersNIGCLZ18} so that it can learn deep contextualized acoustic representations for semi-supervised speech recognition.


Inspired by the MLM proposed in BERT \cite{DBLP:conf/naacl/DevlinCLT19}, recent works \cite{DBLP:conf/icassp/LiuYCHL20,DBLP:journals/corr/abs-2007-06028} have explored using BERT-style objective in speech pre-training. In Mockingjay \cite{DBLP:conf/icassp/LiuYCHL20}, part of input frames are masked to zeros and the pre-trained encoder is required to predict the masked frame from its neighborhood. 
TERA \cite{DBLP:journals/corr/abs-2007-06028} extends Mockingjay by introducing channel alteration and magnitude alteration.

\subsection{Multi-Task Approaches}

PASE \cite{DBLP:conf/interspeech/PascualRSBB19} uses multiple regressors and discriminators to learn a problem-agnostic speech encoder. Another work PASE+ \cite{DBLP:conf/icassp/RavanelliZPSMTB20} improves PASE for robust speech recognition in noisy and reverberant environments by introducing data augmentation, more regression tasks, and a collection of architecture modification.

Our work and PASE both consider combinations of generative and discriminative objectives. However, PASE does not consider speech hierarchy. In our work, different objectives are used to handle signals at different time scales. 

\subsection{Self-Supervised Learning in Other Domains}

Our work is inspired by some self-supervised learning methods in other domains. BERT \cite{DBLP:conf/naacl/DevlinCLT19} is a milestone work for pre-training in NLP. The core of BERT is the MLM, where some input tokens are randomly masked out, and the training objective is to predict the vocabulary ID of the masked word based only on its context. SimCLR \cite{DBLP:conf/icml/ChenK0H20} proposes a simple contrastive learning framework for visual representation learning. It adopts the contrastive loss between augmented views of an image without relying on specialized architecture design or memory bank mechanism. BERT and SimCLR inspired our phoneme-scale and sentence-scale contrastive learning, respectively. 

\section{Multi-Granularity Framework} \label{sec:3}


\subsection{Overview} \label{subsec:3.1}

\begin{figure}[t]
\centering
\includegraphics[width=0.9\columnwidth]{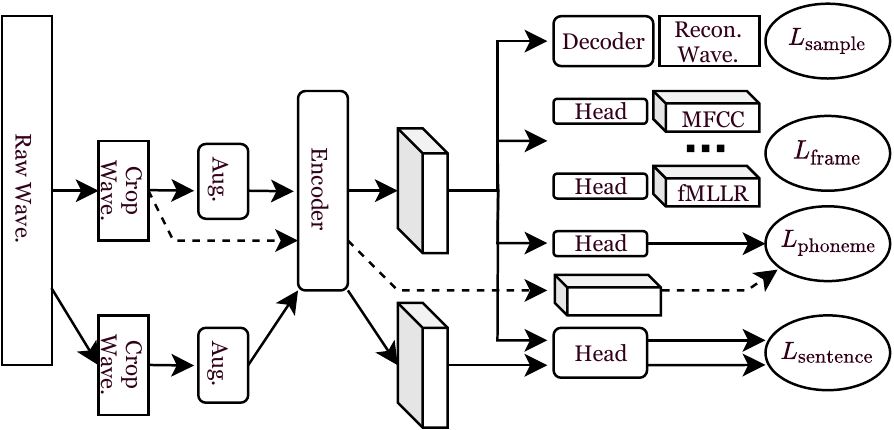}
\caption{A sketch of the multi-granularity framework for self-supervised speech representation learning. Different heads are not weight-sharing. Recon.: Reconstructed. Aug.: Augmentation. Wave.: Waveform. }
\label{fig:framework}
\end{figure}


The main idea behind MGF is to extract the information attached to each speech hierarchy by multi-granularity objectives. At the finest granularity, we adopt generative approaches to reconstruct the original waveform and selected hand-crafted features to extract sample-scale and frame-scale information, respectively. At a coarser granularity, we design a novel continuous masked language model (cMLM) which masks several consequent frames with the typical phoneme length. The model is trained to estimate the feature embedding for one of the masked frames based on the context information. We do not expect the model to recover the exact features, but we hope that the estimated feature for the masked frame is close to the ground truth feature and far away from features of other frames in different phonemes. At the sentence level, learning encourages segments within the same sentence to have close representations and segments across different sentences to have representations that are far apart. 



\subsection{Learning Objectives and Loss Design}
Learning of sample-scale and frame-scale information adopts reconstruction loss, while learning of phoneme-scale and sentence-scale information adopts contrastive loss.

\subsubsection{Sample-Scale Loss} 

As Fig.\ref{fig:framework} shows, a decoder is appended to the base module to reconstruct the original signal from the feature embedding. Let $x$ denote the original signal and $\bar{x}$ denote the reconstructed signal. The sample-scale loss is implemented by the scale-invariant signal-to-distortion ratio (SI-SDR) \cite{DBLP:conf/icassp/RouxWEH19} loss which is formulated as:
\begin{equation} \label{eq:loss1}
L_\text{sample} = -10 \log_{10}(\frac{|\alpha x|^2}{|\alpha x - \bar{x}|^2}) \  \text{for} \ \alpha=\frac{\bar{x}^T x}{|x|^2}. 
\end{equation}
SI-SDR loss is widely used in speech separation \cite{DBLP:conf/ijcai/ZhaoLZZ20} and we empirically find it works better than L1 loss.

\subsubsection{Frame-Scale Loss}

Several heads composed of two convolutional (conv) layers are appended to the base model to generate selected hand-crafted features frame by frame. These features, including log power spectrum (LPS), mel-frequency cepstral coefficient (MFCC), and maximum likelihood linear regression (fMLLR), have been proven effective in speech-related tasks. Following PASE+, we set the context window for LPS and MFCC to both 25ms and 400 ms. L2 loss is used as the optimization target and it is formulated as:
\begin{equation} \label{eq:loss2}
    L_\text{frame} = \underset{u \in \mathcal{U}}{\mathbb{E}}\underset{h \in \mathcal{H}}{\sum}\omega_h||u_h - \hat{u}_h||^2
\end{equation}
where $\mathcal{U}$ is the set of unmasked frames, $\mathcal{H}$ is the set of hand-crafted feature indicators, $\omega_h$ is the weight of $h$ ($h \in \mathcal{H}$), and $u_h, \hat{u}_h$ are the ground-truth and the estimation of feature $h$ for frame $u$, respectively. 

\subsubsection{Phoneme-Scale Loss}
A speech segment that is tens to hundreds of milliseconds long contains distinguishable phonemes. We believe that learning distinguishable high-level semantic information is more important than waveform or feature reconstruction on this time scale. 

Phoneme-scale information is learned by our proposed continuous MLM model and the discriminative learning objective. 
In the vanilla MLM, discrete tokens from a finite dictionary are masked in the language input and then predicted in the output based on their context. However, speech is a continuous signal without token boundaries. 
It is not possible to precisely mask some pre-defined phonemes. To address this challenge, we randomly mask a fixed-length speech segment, and use InfoNCE \cite{DBLP:journals/corr/abs-1807-03748} loss to evaluate the quality of the estimated features of a masked frame. In our implementation, each masked segment has a length of 140ms, and the total length of masked segments does not exceed 20\% of the input speech crop. The masked segment is replaced by non-speech noise. We empirically find that it is a better choice than a segment of zeros or a random speech.

InfoNCE loss directly operates on real-valued feature vectors and is formulated as:
\begin{equation} \label{eq:loss3}
    L_\text{phoneme} = -\underset{v\in \mathcal{V}}{\mathbb{E}}\log\frac{\text{exp}(v^T\hat{v}/\tau_1)}{\sum_{k=1}^{K}\text{exp}(v^T v_k/\tau_1) + \text{exp}(v^T\hat{v}/\tau_1)}
\end{equation}
where $\mathcal{V}$ is the set of masked frames, $v$ is the anchor sample, $\hat{v}$ is the positive sample, and $v_k (k=1,...,K)$ are negative samples. $\tau_1$ is a temperature parameter. 

Note that each sample is the feature representation of a single frame whose duration is 10ms. The anchor sample is drawn from the MGF representation of the masked crop, while the positive sample and negative samples are drawn from the feature representation of the original unmasked crop. In other words, the anchor sample is the estimated feature while positive and negative samples are ground-truth features at the same or different locations, respectively. 




\subsubsection{Sentence-Scale Loss}
Sentence-scale loss focuses on capturing semantic information relevant to a long time scale. We adopt SimCLR framework \cite{DBLP:conf/icml/ChenK0H20}, which was proposed for visual representation learning, to implement the idea. We make some small modifications to SimCLR. First, since cropping is useful only in computing sentence-scale loss, we crop two segments of a sentence before applying other data augmentations. Then, one segment is treated as the original crop for all the other objectives. Second, we use data augmentations specific to speech signal. The augmentations include temporal mask and additive noise. 

The sentence-scale loss is defined as:
\begin{equation} \label{eq:loss4}
    L_\text{sentence} = -\log\frac{\text{exp}(z_i^T z_j/\tau_2)}{\sum_{k=1}^{2N-1}\mathbb{I}_{[k\neq j]}\text{exp}(z_i^T z_k/\tau_2)}
\end{equation}
where $z_i$ is the anchor sample, $z_j$ is the positive sample, $z_k (k \neq j)$ are negative samples, $N$ is batch-size, and $\tau_2$ is a temperature parameter. Each sample is obtained by averaging the MGF representation of all frames within a 2s-long speech crop and passing the averaged feature through a head of two conv layers. The positive sample corresponds to a speech crop from the same sentence as the anchor crop, while negative samples correspond to crops from other sentences. Following SimCLR, we draw negatives from the mini-batch so that there are $2N-2$ negative samples for each anchor.

\subsubsection{Multi-Granularity Objectives}
We train the base encoder by combining multi-granularity objectives. The total loss function is defined as:
\begin{equation} \label{eq:loss5}
    L = \lambda_1 L_\text{sample} + \lambda_2 L_\text{frame} + \lambda_3 L_\text{phoneme} + \lambda_4 L_\text{sentence}
\end{equation}
where $\lambda_i,i=1,2,3,4$ are weights of each loss. We tune each weight in $\{0.03, 0.1, 0.3, 1\}$ independently. 

\subsection{Implementation}

MGF is implemented in PyTorch. The base encoder is composed of three conv layers and six blocks of Transformer. 
The first conv layer has a kernel size of 320 and implements 512 filters with stride 160 and padding 80. The second conv layer has a kernel size of 1 followed by ReLU activation. These two conv layers serve as a stem which transforms the time-domain waveform to a compact feature vector, so that Transformer does not need to handle very long and low-level input. The third conv layer aligns the dimension with the subsequent Transformer. 
In addition to the base encoder, the decoder for sample-level reconstruction also uses four blocks of Transformer. All Transformers share the same parameters with hidden size $d_\text{model}=768$, feed-forward size $d_\text{ff}=3072$, and the number of attention heads $h=12$.

\section{Experiments} \label{sec:4}

\subsection{Experiment Setup}

In most of our experiments, we use the \textit{train-clean-100} subset of the LibriSpeech corpus \cite{DBLP:conf/icassp/PanayotovCPK15} as the pre-training dataset. It contains 100 hours of speech data. We use \textit{dev-clean} subset as the validation dataset for model selection. In the pre-training stage, we use the raw signal only and ignore any human labels such as speaker ID or transcriptions. Some of the MGF objectives rely on data augmentation of additive noise. We use DNS challenge dataset \cite{DBLP:journals/corr/abs-2005-13981}, which contains 70k clips of non-speech noise, for this purpose. 

We evaluate MGF representation in a wide range of downstream tasks. Except for the speech recognition task, linear evaluation is adopted, where the pre-trained model is appended with only one learnable layer. The classification accuracy is used as the evaluation metric. A majority of general-purpose speech pre-training work are evaluated on phoneme and speaker classification tasks. We use these two tasks for ablation studies in addition to the system comparison with methods in the same category. We also evaluated MGF in other downstream tasks to show its generalization capability. 

\textbf{Phoneme classification}: We follow the setup in CPC \cite{DBLP:journals/corr/abs-1807-03748} to use 41 phoneme classes and the \textit{train-clean-100} subset of LibriSpeech for both training and testing. For a fair comparison, we use aligned phone labels and train/test split provided by CPC.

\textbf{Speaker classification}: Following CPC, we first use LibriSpeech \textit{train-clean-100} subset containing 251 speakers for this task. Using the same train/test split provided by CPC, the proposed MGF achieves a classification accuracy of 100\%. In view of the saturated performance, we propose a new setting called \textbf{one-shot speaker classification}. We use the \textit{train-clean-360} subset which contains 921 speakers. For each speaker, we only put one utterance into the training set and sample 20\% from the rest utterances into the test set. This creates 3.2 hours of training data and 72 hours of testing data.

\textbf{Emotion classification}: We use interactive emotional dyadic motion capture (IEMOCAP) dataset \cite{DBLP:journals/lre/BussoBLKMKCLN08} for this task. This corpus consists of five sessions with two speakers in each session. Following the usage in \cite{DBLP:conf/icassp/WuLCLYDMHWLM19}, we evaluate the MGF representation on four emotions, namely \textit{Neutral, Angry, Happy and Sad}. The IEMOCAP dataset contains scripted data and improvised data and we only use the latter. We report results of five-fold cross validation using four sessions as training and the other session as validation and testing.

\textbf{Speech Recognition}: We use Wall Street Journal (WSJ \cite{garofolo1993csr}) dataset for the ASR task. This corpus comprises about 81 hours of transcribed audio data. We train on si284, validate on nov93dev and test on nov92. We use the lexicon-free \cite{DBLP:conf/interspeech/LikhomanenkoSC19} acoustic model and 4-gram KenLM \cite{DBLP:conf/acl/HeafieldPCK13} language model which are implemented by wav2letter++ \cite{DBLP:conf/icassp/PratapHXCKSLC19}. Word error rate (WER) and letter error rate (LER) are used as evaluation metrics. We use the training recipe provided by wav2letter++ and only modifies the input embedding.

For the pre-training and classification downstream tasks, we use Adam optimizer with warm-up to update the model. We use learning rate of \{1e-3, 1e-3, 1e-3, 1e-3, 3e-4\}, warm-up steps of \{10000, 5000, 5000, 5000, 2000\} and batch-size of \{120, 64, 32, 64, 3\} for pre-training, phoneme classification, speaker classification, speaker verification and emotion classification training, respectively. We also exponentially decay the learning rate with exponent of 0.3. We use 4 V100 GPU in both pre-training and downstream task finetuning. The total training epochs for pre-training is set to 300, and more epochs yield slight improvement. For all downstream task finetuning, we set total training epochs to 100 except the experiments in data efficiency.

\subsection{Ablation Study} \label{subsec:4.2}

We first present ablation study of MGF to show the effectiveness of multi-granularity objectives and cMLM. We report accuracies on phoneme classification and one-shot speaker classification tasks.

\subsubsection{Multi-Granularity Framework}

\begin{table}[t]
\centering
\begin{tabular}{ccc}
    \toprule
    Objectives  & \thead{Phoneme Acc} & \thead{One-Shot Speaker Acc}  \\
    \midrule
    MGF (Full)           & 73.4        & 82.7  \\
    -Sample              & 73.3 (-0.1) & 82.2 (-0.5)\\
    -Frame               & 69.2 (-4.2) & 79.9 (-2.8)\\
    -Phoneme             & 66.1 (-7.3) & 81.3 (-1.4)\\
    -Sentence            & 73.4 (-0.0) & 81.8 (-0.9)\\
\bottomrule
\end{tabular}
\caption{Classification accuracies on phoneme classification and one-shot speaker classification using MGF representation. First row are results of full model. Following rows report results when discarding each objective. Numbers in the parenthesis emphasize accuracy drop.}
\label{tab:1}
\end{table}

We study whether all four objectives at different time scales contribute to the final accuracy of MGF, and assess their respective importance on two target problems. We trained MGF four times, discarding one of the four loss objectives at a time. Results are presented in Table \ref{tab:1}. The first row presents the results of the full model and the other rows present the results when a certain objective is discarded.

The first finding is that every objective matters. Discarding any objective leads to notable accuracy drop in at least one task. Second, while some objectives have general impact on two tasks, others turn out to be more task-oriented. For example, sample-scale loss and frame-scale loss are generally helpful. This is consistent with our design as these two losses learn low-level problem-agnostic information. The frame-scale loss is specially important as it injects human prior knowledge into the model. The other two losses, however, are more task-dependent. Phoneme-scale loss has a remarkable impact on phoneme classification task (+27.4\% in relative error) and sentence-scale loss only contribute to one-shot speaker classification task. It is worth noting that phoneme-scale loss also contributes a lot to the one-shot speaker classification task. This is caused by the sampling strategy used in cMLM, which will be described next. 

\subsubsection{cMLM}

\begin{table}[t]
\centering
\begin{tabular}{ccc}
    \toprule
    Approach & \thead{Phoneme Acc } & \thead{One-Shot Speaker Acc } \\
    \midrule
    Discriminative & 73.4 & 82.7\\
    Generative     & 66.1 & 81.3\\
\bottomrule
\end{tabular}
\caption{Comparison of discriminative and generative approaches for phoneme-scale training.}
\label{tab:2}
\end{table}


MGF uses cMLM and adopts InfoNCE loss for phoneme-scale target. Previous work \cite{DBLP:conf/icassp/LiuYCHL20} has investigated a similar masking approach but has used a generative approach with reconstruction loss at a similar time scale. We believe that our proposed discriminative approach with InfoNCE loss is more suitable for this time scale. To validate this, we implemented a generative approach which calculates L1 loss between the predicted frame and the ground-truth frame in the masked segment. The two rows in Table \ref{tab:2} show the results of using discriminative (InfoNCE) loss and generative (L1) loss, respectively. The model trained with discriminative approach achieves 7.3\% accuracy gain on phoneme classification and 1.4\% accuracy gain on one-shot speaker classification, compared with the model trained with generative approach. Coincidentally, the speaker classification accuracy achieved by the generative approach (81.3\%) is the same as MGF without phoneme-scale objective. In other words, using generative approach for phoneme-scale learning does not help the speaker classification task at all. 

It is worth noting that the sample strategy in cMLM has notable impact on the performance of MGF representation. There are two options to choose negative samples in cMLM: from a different sentence or from the same sentence as the positive sample. Experimental results show that sampling from a \textit{different sentence} leads to a better performance (1.2\% and 0.9\% accuracy gain on the two tasks, respectively). The reason is that a richer vocabulary is helpful for the model to learn more discriminative features. In addition, the \textit{different-sentence} sample strategy allows the model to learn features that are able to discriminate speakers.

\subsection{Comparison with General-Purpose Self-Supervised Speech Pre-Training Methods}


\begin{table}
\centering
\begin{tabular}{ccc}
    \toprule
    Method &  Phoneme Acc & Speaker Acc  \\
    \midrule
    MFCC                &39.7 & 17.6\\
    \midrule
    CPC                 &65.5 & 97.4\\
    Mockingjay          &64.3 & 96.1\\
    TERA-base (3xT)     &65.1 & 99.2\\
    TERA-medium (6xT)   &65.9 & -\\

    MGF (6xT)           &73.4 & 100.0\\
\bottomrule
\end{tabular}
\caption{Comparison of self-supervised speech representation learning methods on Librispeech phoneme and speaker classification tasks under linear evaluation. $n$xT denotes $n$ Transformer blocks.}
\label{tab:5}
\end{table}

We compare MGF with CPC \cite{DBLP:journals/corr/abs-1807-03748}, Mockingjay \cite{DBLP:conf/icassp/LiuYCHL20}, and TERA \cite{DBLP:journals/corr/abs-2007-06028} on phoneme classification and speaker classification tasks. All the systems use the same setup as specified in CPC. Results are shown in Table \ref{tab:5}. 

CPC \cite{DBLP:journals/corr/abs-1807-03748} is a discriminative approach. We have pointed out earlier that it is not appropriate in CPC to distinguish positive and negative samples only based on the distance to the anchor. In contrast, MGF uses a masked model. When the input crop is masked or unmasked, features at the same location always form a positive pair. 
In addition, MGF is a multi-granularity framework and uses bidirectional model. These advances explain why MGF outperforms CPC by a large margin on both tasks. Note that the evaluated MGF model has 12G FLOPs, which is a bit heavier than CPC which has 9.7G FLOPs. 

Mockingjay \cite{DBLP:conf/icassp/LiuYCHL20} and TERA \cite{DBLP:journals/corr/abs-2007-06028} are generative approaches which try to predict acoustic frames from its manipulated version. Mockingjay only uses temporal alteration and TERA extends it by adding channel alteration and magnitude alteration. In the previous section, we have tried similar generative loss and find that discriminative loss works better. Table \ref{tab:5} shows that MGF gets 9.1\%/8.3\% accuracy gain on phoneme classification and 3.9\%/0.8\% accuracy gain on speaker classification over Mockingjay and TERA, respectively.


\subsection{Evaluation on More Tasks}

\subsubsection{One-Shot Speaker Classification}

We additionally evaluate MGF representation against two well-known feature embeddings, namely d-vector \cite{DBLP:conf/icassp/WanWPL18} and x-vector \cite{DBLP:conf/icassp/SnyderGSPK18}, in one-shot speaker classification task. d-vector is learned via a generalized end-to-end loss, which is similar to the triplet loss. x-vector is learned via a speaker recognition task using a time-delay DNN. 
In particular, x-vector is recognized as the state-of-the-art embedding for speaker classification tasks. d-vector and x-vector are both pre-trained on Switchborad \cite{DBLP:conf/icassp/GodfreyHM92} and NIST SREs \cite{DBLP:journals/speech/DoddingtonPMR00} datasets and x-vector is implemented officially via Kaldi's V2 recipe \cite{povey2011kaldi}. 

We use the same linear evaluation protocol for all the methods to ensure a fair comparison. d-vector achieves 77.8\% accuracy and x-vector achieves 79.6\% accuracy. As a comparison, MGF achieves 82.7\% accuracy which reduces the relative error by 15.2\% compared with the best counterpart. 

\subsubsection{Speech Recognition}

\begin{table}[t]
\centering
\begin{tabular}{ccccc}
    \toprule
    Method & \multicolumn{2}{c}{nov93dev} & \multicolumn{2}{c}{nov92} \\
     & \thead{LER} & \thead{WER} & \thead{LER} & \thead{WER}\\
    \midrule
    Baseline        & 3.50  & 8.57  & 2.09 & 5.42\\
    wav2vec-large   & 2.91  & 7.24  & 1.64 & 4.48\\
    MGF-960         & 3.07  & 7.58  & 1.78 & 4.87\\
\bottomrule
\end{tabular}
\caption{WSJ speech recognition results.}
\label{tab:4}
\end{table}

We evaluate the performance of an ASR system built on top of MGF representation. The baseline method uses 80 log-mel filterbank coefficients with a 25ms sliding window and 10ms stride. We also compare MGF with a well-known ASR-oriented self-supervised method wav2vec \cite{DBLP:conf/interspeech/SchneiderBCA19}. We use their released wav2vec-large checkpoint\footnote{https://github.com/pytorch/fairseq/tree/master/examples/wav2vec}. As wav2vec is pre-trained with the entire 960 hours of LibriSpeech training set, we train the model MGF-960 with the same amount of training data to ensure a fair comparison. MGF-960 has the same architecture and model size as our base model. As shown in Table \ref{tab:4}, MGF-960 achieves significantly lower WER than the baseline and comparable WER as wav2vec-large. This means, without bells and whistles, our general-purpose speech representation can benefit the speech recognition task.

\subsubsection{Emotion Classification}

\begin{table}
\centering
\begin{tabular}{cccc}
    \toprule
    Method & Modality & Emotion Acc   \\
    \midrule
    M3ER            & AVT & 82.7\\
    CNN LSTM        & A   & 68.8\\
    CNN GRU-SeqCap  & A   & 72.7\\
    \midrule
    MGF-Scratch     & A   & 64.1 \\
    MGF-Fixed       & A   & 71.2\\
    MGF-Finetune    & A   & 73.1\\
\bottomrule
\end{tabular}
\caption{IEMOCAP emotion classification accuracies of different methods. A, V, T are shorts for audio, video and text, respectively.}
\label{tab:6}
\end{table}

We present experimental results of emotion classification on IEMOCAP dataset. We want to use this task to evaluate the adaptation capability of MGF representation. Since this task is never evaluated by previous self-supervised representation learning methods, we compare MGF with state-of-the-art supervised methods instead. M3ER \cite{DBLP:conf/aaai/MittalBCBM20} uses text, audio and video to predict speaker's emotion. CNN GRU-SeqCap \cite{DBLP:conf/icassp/WuLCLYDMHWLM19}, CNN LSTM \cite{DBLP:conf/interspeech/SattRH17}, and our MGF only use audio. We set three different settings for MGF. MGF-Scratch does not have pre-training stage while MGF-Fixed and MGF-Finetune are pre-trained. The base encoder of MGF is not trainable in \`fixed" and trainable in \`Finetune". As shown in Table \ref{tab:6}, MGF-Scratch does not work well but MGF-Fixed and MGF-Finetune both achieve high accuracy, showing that pre-training does do a lot help. MGF-Fintune even creates a new SOTA among audio-only methods.


\subsection{Data Efficiency}

\begin{figure}[t]
    \centering
\begin{subfigure}[b]{0.45\textwidth}
    \begin{tikzpicture}[scale=0.85]
        \begin{semilogxaxis}[
            xlabel=Data Usage (\%),
            ylabel=Phoneme Accuracy (\%),
            height=0.7\textwidth,
            width=1.1\columnwidth,
            compat=1.3,
            legend pos=south east,
            title style={font=\large},
            label style={font=\large}]

        \addplot[dashed, smooth,mark=square*, color=blue] plot coordinates {
            (0.1, 34.87)
            (0.5, 55.35)
            (1, 64.52)
            (10,83.15)
            (25,86.58)
            (50,88.40)
            (75,89.23)
            (100,89.72)
        };
        \addlegendentry{train-from-scratch}
        
        \addplot[dash dot, smooth,mark=triangle*, color=red] plot coordinates {
            (0.1, 72.25)
            (0.5, 78.23)
            (1, 80.11)
            (10,85.63)
            (25,87.83)
            (50,89.20)
            (75,89.80)
            (100,90.24)
        };
        \addlegendentry{pre-trained }

        \end{semilogxaxis}
    \end{tikzpicture}
    \end{subfigure}
    \caption{Comparison of representations with phoneme classification accuracy across different amount of labeled data.}
    \label{fig:2}
\end{figure}
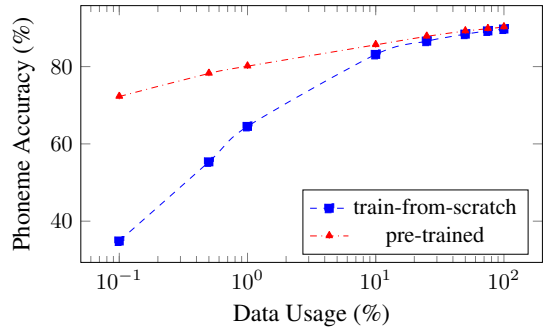

Last but not least, we show how pre-training could help in low-resource scenarios where human labels are scarce. We use LibriSpeech phoneme classification task and reduce the labeled data usage from 100\% to 0.1\%. The performance of MGF in different settings are plotted in Figure \ref{fig:2}. To compare fairly with the model without pre-training, we open up the entire model for fine-tuning. We find that pre-trained MGF outperforms its train-from-scratch counterpart by a large margin when the length of labeled data is less than one hour. In an extreme low-resource scenario where only six minutes of labeled data are available, the pre-trained model still achieves a reasonably good performance of 72.3\% phoneme accuracy while the train-from-scratch model is only able to achieve 34.9\% phoneme accuracy.

\section{Conclusion and Future Work} \label{sec:5}
We have proposed a multi-granularity framework for self-supervised speech representation learning. By taking the speech hierarchy into consideration, MGF achieves top performance among existing speech pre-training methods on a collection of speech tasks. Comprehensive ablation studies have been carried out to demonstrate the effectiveness of our design in MGF. In the future, we plan to expand this multi-granularity self-supervised framework to the image domain, which may benefit tasks that demand multi-scale features.


\bibliographystyle{named}
{\small
\bibliography{ijcai21}
}
\end{document}